# Detailed crystallographic analysis of the ice VI to ice XV hydrogen ordering phase transition

Christoph G. Salzmann,*[a] Ben Slater,[a] Paolo G. Radaelli,[b] John L. Finney,[c] Jacob J. Shephard,[a] Martin Rosillo-Lopez[a] and James Hindley[a]

[a] Department of Chemistry, University College London, 20 Gordon Street, London WC1H 0AJ, United Kingdom; E-mail: c.salzmann@ucl.ac.uk
[b] Department of Physics, University of Oxford, Parks Road, Oxford OX1 3PU, United Kingdom
[c] Department of Physics and Astronomy, University College London, Gower Street, London WC1E 6BT, United Kingdom

**Abstract**

The $D_2O$ ice VI to ice XV hydrogen ordering phase transition at ambient pressure is investigated in detail with neutron diffraction. The lattice constants are found to be sensitive indicators for hydrogen ordering. The *a* and *b* lattice constants contract whereas a pronounced expansion in *c* is found upon hydrogen ordering. Overall, the hydrogen ordering transition goes along with a small increase in volume which explains why the phase transition is more difficult to observe upon cooling under pressure. Slow-cooling ice VI at 1.4 GPa gives essentially fully hydrogen-disordered ice VI. Consistent with earlier studies, the ice XV obtained after slow-cooling at ambient pressure is best described with *P*-1 space group symmetry. Using a new computational approach, we achieve the atomistic reconstruction of a supercell structure that is consistent with the average partially ordered structure derived from Rietveld refinements. This shows that *C*-type networks are most prevalent in ice XV but other structural motifs outside of the classifications of the fully hydrogen-ordered networks are identified as well. The recently proposed *Pmmn* structural model for ice XV is found to be incompatible with our diffraction data and we argue that only structural models that are capable of describing full hydrogen order should be used.

**Introduction**

The ice VI high-pressure phase was discovered by Bridgman when he explored the phase diagram of $H_2O$ up to 2 GPa.[1] The ice V/VI/liquid triple point is located at 0.632 GPa and 0.1°C which means that ice VI was the first of the high-pressure ices to display a melting point higher than the 'ordinary' ice I$h$ at ambient pressure. Referring to the newly discovered ice VI, the Los Angeles Evening Herald reported on the 23$^{rd}$ of August 1912 that Bridgman had succeeded in making "hot ice" at a temperature of 78°C. Requests from industry regarding the use of ice VI in refrigeration were swiftly dismissed by Bridgman.[2] The crystal structure of ice VI was solved by Kamb from X-ray diffraction patterns of ice VI recovered at ambient pressure under liquid nitrogen.[3] Figure 1(a) shows the tetragonal unit cell of ice VI which contains ten water molecules and displays $P4_2/nmc$ space group symmetry. Ice VI consists of two interpenetrating yet independent hydrogen-bonded networks, and has therefore been described as a "self-clathrate".[3] The individual networks consist of hexameric units of water molecules which have the same structure as the $(H_2O)_6$ 'cage-like' cluster in the gas phase (*cf.* Figure 1(b)).[4,5] In ice VI, these clusters share corners in the *c* direction and they are hydrogen-bonded to one another in the *a* and *b* directions. The resulting three-dimensional networks are remarkable in the sense that they contain only four and eight-membered rings.[6] Depending on the positions of the water molecules within the hexameric units we classify them as either *apex* or *waist* molecules as shown in Figure 1(b).

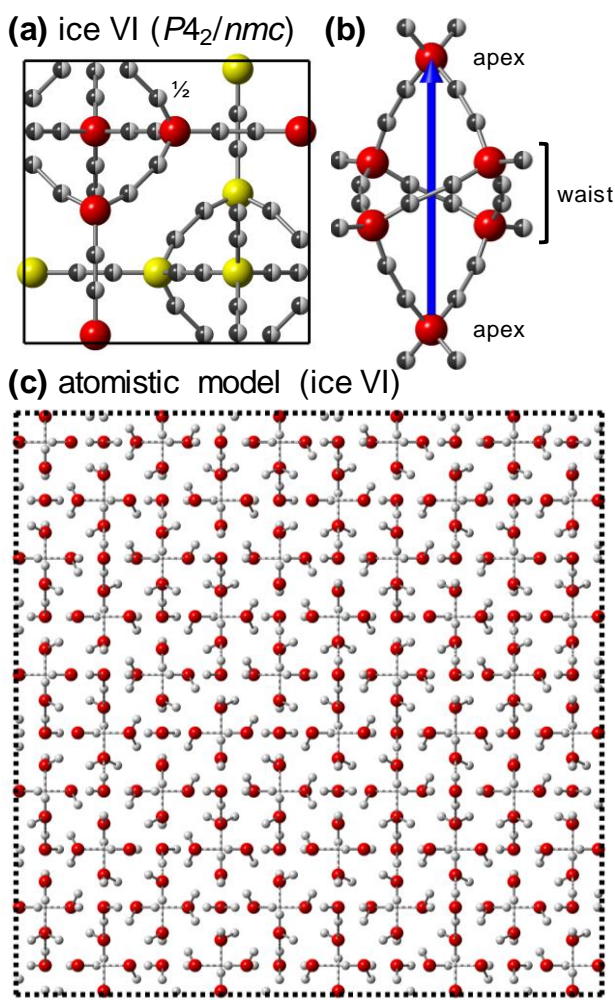

***Fig. 1*** *(a) Unit-cell projection of ice VI representing the average P4$_2$/nmc structure (origin choice 2) viewed along the [001] direction. The oxygen atoms of the two different networks are shown in red and yellow, respectively. The hydrogen positions are half occupied. (b) The hexameric unit of a single network viewed along [110]. The four-fold rotoinversion axis is shown as a blue vector. (c) Atomistic model of a 5x5x6 supercell consistent with fully hydrogen-disordered ice VI. The maximal (average) deviation from ½ occupancy is 0.013 (0.005). To prevent atomic overlaps, only atoms in the 0 < z < 1/6 range are shown. The dashed lines of the supercell indicate that the periodic boundary conditions would not be required in ice VI. Covalent O-H bonds are indicated by solid and O···H hydrogen bonds by dashed lines.*

The *P*4$_2$/*nmc* space group symmetry implies that the hydrogen-bonded water molecules in ice VI are fully orientationally disordered which gives rise to an *average* structure with half-occupied hydrogen sites as shown in Figure 1(a). This structure was confirmed by Kuhs *et al.* who performed neutron diffraction on ice VI samples under *in-situ* pressure conditions.[7, 8] The hydrogen-disordered nature of ice VI in its region of thermodynamic stability is also consistent with dielectric[9] and calorimetric measurements.[10] Figure 1(c) shows an atomistic supercell representative of hydrogen-disordered ice VI which was generated by our *RandomIce* program (*cf.* Experimental and Computational Methods section). The hydrogen disorder can be seen, for example, from the random orientations of the *apex* water molecules which appear as either straight or at a right angle in the projection shown in Figure 1(c). It is important to stress that the crystallographic 'trick' of using half-occupied hydrogen sites for describing the average structures of hydrogen-disordered ices gives rise to symmetry that does not reflect the local atomic structure. In fact, the hydrogen disorder destroys the translational symmetry entirely and, strictly speaking, the hydrogen-disordered ices should be classified as non-crystalline materials.

There were early reports of hydrogen ordering of ice VI at low temperatures. In fact, Kamb mentioned the observation of a 001 reflection in the X-ray diffraction of ice VI at ambient pressure and 98 K which is inconsistent with one of the reflection conditions of *P*4$_2$/*nmc* (00*l*: *l* = 2*n*).[3] He later assigned *Pmmn* space group symmetry to what he called ice VI'.[11] *Pmmn* allows only partial hydrogen order and implies anti-ferroelectric hydrogen ordering. This was, however, not reproduced in later neutron diffraction work.[7, 8] Weak hydrogen ordering upon cooling ice VI under pressure has been suggested from dielectric,[12, 13] thermal conductivity[14] and thermal expansion measurements.[15] In contrast to Kamb's ice VI',[11] the dielectric measurements implied ferroelectric ordering of ice VI upon cooling under pressure.[12, 13] Raman and FT-IR spectroscopy of ice VI at low temperatures did not identify any signs of hydrogen order,[16, 17] even when the sample was doped with potassium hydroxide[18] which has been found to promote hydrogen ordering in ice I*h*.[19]

Using neutron diffraction, Salzmann *et al.* showed that doping with hydrochloric acid (HCl) enables the phase transition of ice VI to its hydrogen-ordered counterpart ice XV to take place upon cooling.[20] Slow-cooling at ambient pressure was found to deliver a more hydrogen-ordered sample than at 0.9 GPa. The neutron diffraction patterns of ice XV showed several new Bragg peaks indicating lower space group symmetry compared to $P4_2/nmc$ ice VI. No supercell peaks were identified which would indicate an increase in the size of the unit cell. Depending on the symmetry relationship between the two fully hydrogen-ordered networks space groups $P$-1, $P2_1$, $Pc$ and $Cc$ were suggested. $P$-1 gives overall antiferroelectric structures whereas the other space groups yield ferroelectric structures with different degrees of overall polarity. The individual networks can adopt three different chiral and polar structures labelled *A*, *B* and *C* in ref. 20. The polarity of the *C*-type network was found to be much greater compared to the *A* and *B* networks. On the basis of Rietveld refinements it was concluded that ice XV contains the highly polar *C*-type networks and is antiferroelectric overall with *P*-1 space group symmetry. The situation is therefore similar compared to the hydrogen-ordered ice VIII which contains two hydrogen-ordered cubic ice networks of highest possible polarity[21] and is antiferroelectric overall.[22]

In contrast to the experimental structure, DFT calculations suggested that the lowest energy structure contains *A*-type networks and is ferroelectric with *Cc* space group symmetry.[23, 24] However, a later study using fragment-based 2$^{nd}$ order perturbation and coupled cluster theory identified the antiferroelectric *P*-1 structure with *C*-type networks as the lowest energy structure.[25] This result was subsequently contested using fully periodic 2$^{nd}$ order Møller-Plesset perturbation theory and random phase approximation which reconfirmed the original *Cc* structure with *A*-type networks.[26] However, in this study it was also shown that the energies of the various possible ferroelectric structures are influenced by the dielectric properties of the surrounding medium. A Raman spectroscopic analysis of ice XV showed that the experimental spectra are inconsistent with *Cc* space group symmetry.[27] This space group is also incompatible with the observed 003 reflection of ice XV (0$kl$: $k + l = 2n$).

A recent calorimetric study showed that the kinetics of the ice VI to ice XV phase transition are complex with a fast process taking place first upon cooling ice VI followed by a slower process which manifests as a tail to the initial sharp exotherm.[28] It was suggested that the initial stages of the hydrogen ordering processes could be ferroelectric which is equivalent to saying that the two independent networks may hydrogen order to some extent independently at first. The calorimetric study also showed that it is difficult to achieve complete hydrogen order in ice XV. The maximal loss of configurational entropy during the $H_2O$ ice VI to ice XV was about 50% of the Pauling entropy, $R \ln(3/2)$, which is the value expected for the transition from complete hydrogen disorder to a fully hydrogen-ordered structure.[29] In line with Kamb's original suggestion,[11] Komatsu and co-workers recently suggested that partially hydrogen-ordered ice XV has *Pmmn* space group symmetry.[30]

Research into hydrogen ordering phase transitions is important with respect to the still poorly understood vibrational spectroscopy of condensed water phases.[27] The hydrogen-ordered phases are the only ones for which the spectroscopic selection rules are firmly defined.[27, 31] Knowledge about the low-temperature phase transitions of ice is also needed in geology[32] and the experimentally obtained crystal structures of the hydrogen-ordered ices have been used extensively to benchmark the computational models of $H_2O$.[23-26, 33-35] Furthermore, experimental work on hydrogen ordering has given new insights into the complex nature of defect dynamics in ice.[28, 36-38] Most recently, this knowledge could be successfully applied to the amorphous ices, and it was shown that the glass transitions of low- and high-density amorphous ice are governed by molecular reorientation processes.[39]

Here we present a detailed study on the ice VI to ice XV phase transition using neutron diffraction in order to gain further insights into the characteristics and mechanistic details of the phase transition. The crystallographic details of the phase transition are presented, and particular emphasis is placed on describing partially hydrogen-ordered states. We compare the experimentally observed changes in lattice constants during the hydrogen-ordering phase transition with the lattice constants from DFT calculations, and test the suggested *Pmmn* structural model.

**Experimental and Computational Methods**

*Preparations of DCl-doped ice VI / XV samples*

An indium cup was placed inside a piston-cylinder setup and the entire assembly was cooled with liquid nitrogen from the outside. 600 μl of a 0.01 M DCl solution in $D_2O$ were then quickly pipetted into the indium cup. The ice sample was then pressurised to 1.0 GPa, heated isobarically to 250 K and quenched at ~40 K min$^{-1}$ with liquid nitrogen before releasing the pressure. The sample was then pushed outside the piston cylinder under liquid nitrogen and freed from the indium with a sharp blade. An additional sample was prepared by heating at 1.4 GPa to 250 K and slowly cooled at 0.5 K min$^{-1}$ to 77 K. Using a porcelain pestle and mortar under liquid nitrogen the ice samples were ground to give a powder and transferred into tubular vanadium sample cans for neutron diffraction.

*Neutron diffraction measurements and data analysis*

The filled vanadium cans were mounted onto cryostat sticks under liquid nitrogen and quickly lowered into a precooled AS Scientific "Orange" cryostat on the POLARIS beamline at the ISIS spallation neutron and muon source (Rutherford Appleton Laboratory, UK). The powder neutron diffraction data were analysed with the *GSAS* software[40] by refining structural models to fit the diffraction data from detector banks 3, 4 and 5. Fully hydrogen-ordered structures were used as the starting structures for the Rietveld refinements. To enforce the ice rules, linear constraints were used to ensure that the sums of the fractional occupancies of two hydrogen sites along the hydrogen bonds equal one and chemical composition restraints were implemented to keep the sums over the occupancies of the four hydrogen sites closest to an oxygen atom close to two.

*Preparation of hydrogen-disordered and partially ordered supercell structures*

Our *RandomIce* computer program changes the orientations of water molecules in supercell structures so that the average structure is as close as possible to user-defined fractional occupancies of the hydrogen sites within the small unit cell that is used to describe the average structure. The starting structures for *RandomIce* are *P*1 supercells of a hydrogen-ordered structure containing the atomic positions of the oxygen atoms as well as the occupied and empty hydrogen positions. First the program builds up a connectivity list of the hydrogen-bonded network by evaluating the O-H and H-H pair-distribution functions with user-defined distance limits. The connectivity list is verified with respect to obeying the ice rules. This means that each oxygen atom must have four associated hydrogen positions (two occupied and two empty), and two hydrogen positions must be found along each of the hydrogen bonds (one occupied and one empty). The program then randomly selects one of the oxygen atoms as well as one of the two occupied hydrogen sites associated with that oxygen atom. The occupancy of the hydrogen site is then set to 0 and the occupancy of the empty hydrogen site along the hydrogen bond in question is set to 1. Structurally speaking, this process creates an $H_3O^+$ point defect next to $OH^-$. The rule is then that recombination must not take place but instead, one of the other occupied hydrogen sites of the $H_3O^+$, which is randomly selected, moves on to the next water molecule. This process is repeated until the travelling $H_3O^+$ defect encounters the original $OH^-$ site. Once this has happened, the entire structure consists of only $H_2O$ molecules which have differently oriented water molecules compared to the starting structure. For the new structure, the fractional occupancies of the hydrogen sites of the average structure are calculated and compared with the target values. If the maximal deviation from the target occupancies is below a threshold-value defined by the user, then the structure is stored and the program terminates. If this is not the case, then the program moves on to perform another defect-migration step *etc*. This version of *RandomIce* works very well for preparing fully hydrogen-disordered structures with target occupancies of 0.5. Yet, the preparation of partially hydrogen-ordered structures proved difficult since these structures are statistically speaking less likely encountered when the defect migration is purely random.

In an update to the program, the rule was implemented that out of the two hydrogen atoms that can move on to the next molecule the one is selected that has the lower target occupancy. In addition, a Monte-Carlo type parameter was introduced that defines the fraction of defect moves for which this rule is overridden. This parameter therefore allows choosing a type of defect migration that is in between completely random and strictly-guided by the target occupancies. Depending on the values of the target occupancies, it was necessary to find the best value for the Monte-Carlo parameter by trail-and-error before finally running the program for an extended period of time to obtain a structure closest to the target occupancies. Using a 5x5x6 supercell of ice VI and a computer with a 1.8 GHz i5 processor allowed *RandomIce* to generate and evaluate several hundred million structures in an overnight run.

*Density functional theory (DFT) calculations*

The DFT calculations reported herein were performed with the CP2K program[41] based on the Gaussian and Plane-Waves method.[42, 43] Goedecker-Teter-Hutter pseudopotentials are used to represent the core electrons. A plane wave cut-off of 1200 Ry was used to evaluate the equilibrium structure and density of each of the hydrogen-ordered structures considered with the PBE functional under an isotropic external pressure of 1 GPa. A TZVP basis set was used throughout which has been found to be more than sufficient to capture the energy differences between the different structures accurately. The computational setup used here is almost identical to that used in our previous study on ice XV.[26] Calculations were performed on 2x2x2 supercells containing 80 water molecules such that the supercell lattice constants exceed 10 Å, which is necessary since at the time the calculations were preformed, only Gamma point calculations were possible in CP2K.

From previous studies on a wide range of hydrogen-ordered phases[26, 44, 45] we have found that although the absolute density is sensitive to the functional used, the inclusion and formulation of the van der Waals potential, the relative difference in density for different hydrogen-ordered configurations of a given phase, and the ratio of the lattice constants is not sensitive to the formulation of the functional used.

**Results and discussion**

Figure 2 shows the powder neutron diffraction data recorded at ambient pressure upon heating a pressure-quenched DCl-doped $D_2O$ sample. The initial pattern is consistent with hydrogen-disordered ice VI since it does not display the characteristic additional Bragg peaks of ice XV.[20] These appear upon heating at 2.13, 2.07, 2.02, 1.94, 1.78 and 1.64 Å above 110 K, and disappear again at about 130 K. This indicates a sequence of phase transitions from ice VI to ice XV and back to ice VI upon heating which has also been observed in a recent calorimetric study for an HCl-doped pressure-quenched $H_2O$ sample but not the corresponding $D_2O$ sample.[28] The heating rate in ref. 28 was 10 K min$^{-1}$ which is significantly higher compared to the heating rate of 0.4 K min$^{-1}$ used here. The neutron diffraction data shows that, given more time during heating, the $D_2O$ pressure-quenched sample also undergoes hydrogen ordering just as previously observed for the corresponding $H_2O$ sample.[28]

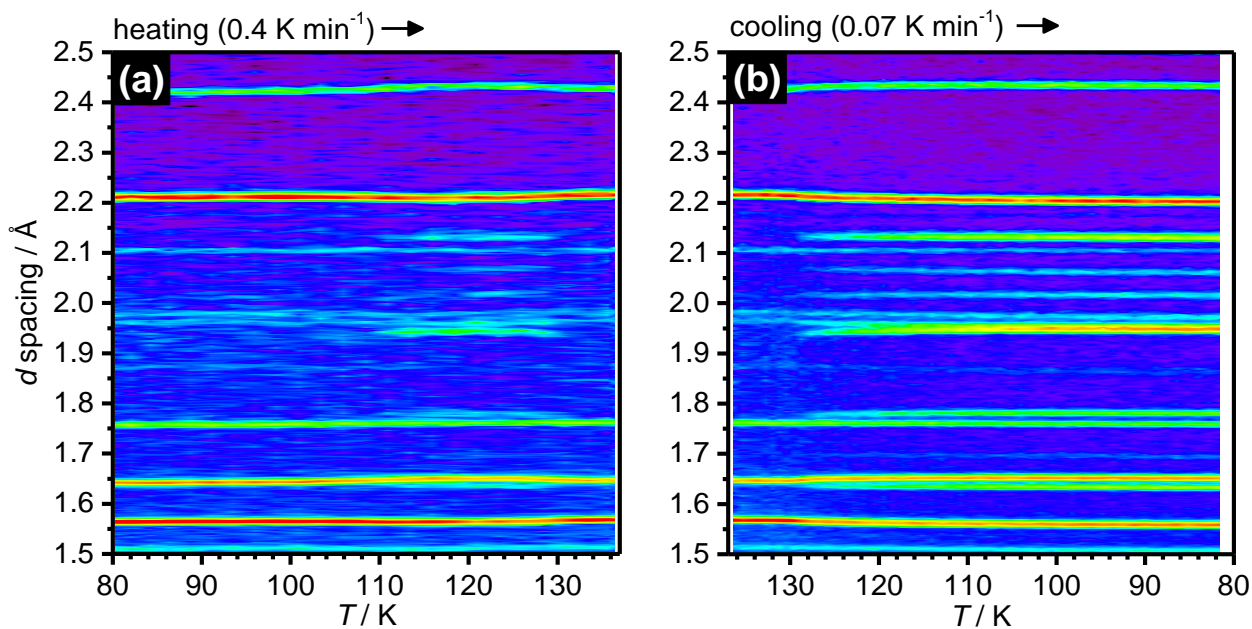

*Fig. 2 Powder neutron diffraction data of pressure-quenched 0.01 mol L$^{-1}$ DCl-doped ice VI recorded (a) upon heating at 0.4 K min$^{-1}$ and (b) subsequent cooling at 0.07 K min$^{-1}$ at ambient pressure. Diffraction intensity at ~2.1 Å is due to vanadium from the sample container.*

The diffraction pattern of the hydrogen-disordered ice VI at 135 K is shown as pattern (1) in Figure 3. A Rietveld refinement showed that there are no detectable signs of partial hydrogen order in ice VI as has been previously observed for the hydrogen-disordered ices III and V.[46] Given the fact that the reorientation dynamics are fast in the DCl-doped sample the fully hydrogen-disordered state of ice VI appears to be the equilibrium state at 135 K and ambient pressure.

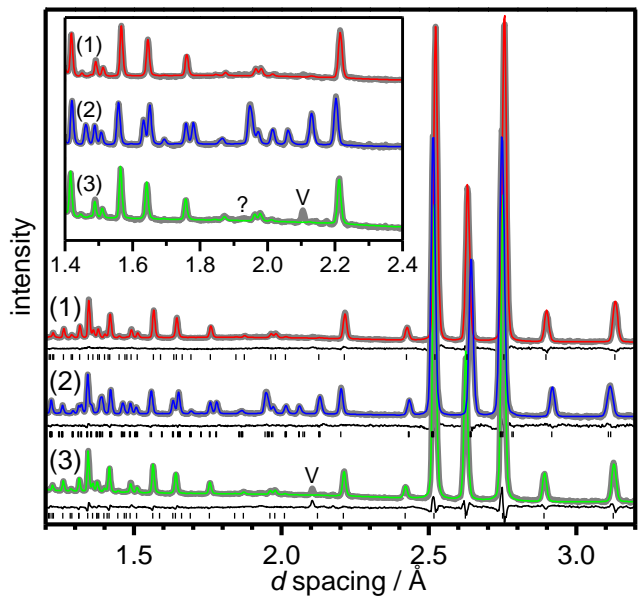

*Fig. 3* Powder neutron diffraction patterns at ambient pressure of DCl-doped (1) ice VI at 135 K, (2) ice XV at 80 K after cooling at 0.07 K min$^{-1}$ at ambient pressure and (3) ice VI at 80 K after cooling at 0.5 K min$^{-1}$ at 1.4 GPa. The experimental data is shown as thick grey lines and the thinner coloured lines are the Rietveld fits. Thin black lines show the differences between the experimental and fitted data. Tickmarks indicate the expected positions of Bragg peaks. The inset shows the region in which the most pronounced differences between ice VI and ice XV have been found. Diffraction intensity at ~2.1 Å in pattern (3) is due to vanadium from the sample container.

Similar to the procedure in ref. 20, the hydrogen-disordered sample was slow-cooled at 0.07 K min$^{-1}$ from 135 K in a next step. The hydrogen-ordering phase transition sets in at about 130 K and the Bragg peaks characteristic for ice XV start to grow in intensity as the temperature decreases. Judging from the intensities of the ice XV peaks at low temperatures, a more ordered state is reached compared to the 'transient' ice XV obtained previously upon heating. Pattern (2) in Figure 3 shows a Rietveld fit to the diffraction data at 80 K using the *P*-1 model of ice XV described in ref. 20.

Changes in the lattice constants have previously been found to be sensitive indicators for hydrogen-ordering processes in ices V/XIII and XII/XIV.[47, 48] Figure 4 shows how the lattice constants of the pressure-quenched sample change upon heating and subsequent cooling. The lattice constants were obtained from Rietveld refinements using the triclinic unit cell of ice XV. The deviations of the triclinic angles from 90° were found to not be significant and have been omitted from Figure 4. The 'transient' hydrogen ordering upon heating the pressure-quenched sample is clearly seen by contractions in the *a* and *b* lattice constants as well as an expansion in *c* before the sample reverts back to ice VI. As expected, these changes become more pronounced as the heating rate was lower from 0.4 K min$^{-1}$ to 0.05 K min$^{-1}$ and the sample is given more time to undergo hydrogen ordering at a given temperature. The *a* lattice constant is more sensitive to hydrogen ordering than *b*. But it is emphasised that this depends entirely on how the structure of ice XV is defined since the two lattice constants are, in principle, interchangeable when the structure of ice XV is derived from the higher symmetry tetragonal ice VI. Above 130 K, the *a* and *b* lattice constants take the same values which is consistent with the tetragonal unit cell of the hydrogen-disordered ice VI.

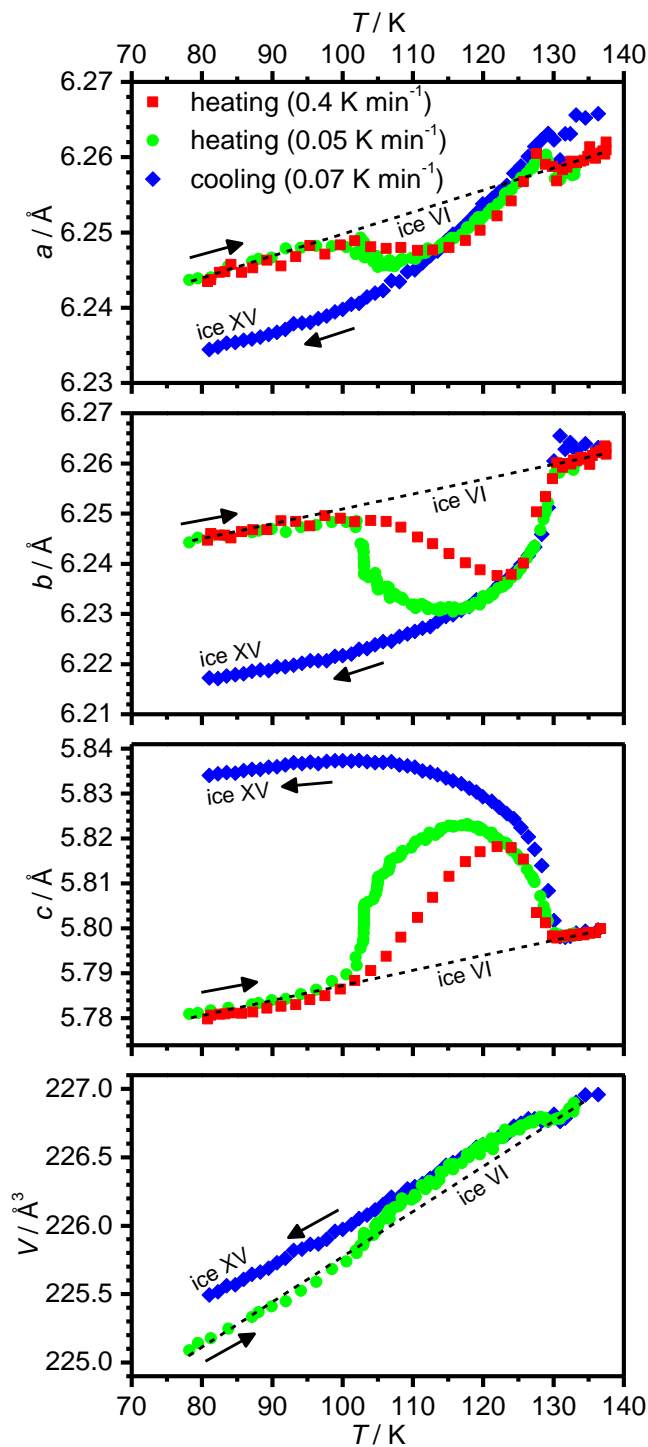

***Fig. 4*** *Changes in the a, b and c lattice constants as well as the unit-cell volume upon heating pressure-quenched 0.01 mol L$^{-1}$ DCl-doped ice VI at 0.4 K min$^{-1}$ (red squares) and 0.05 K min$^{-1}$ (green circles), and upon subsequent cooling at 0.07 K min$^{-1}$ (blue diamonds). The dashed lines indicate the expected trends for ice VI without hydrogen ordering.*

Upon cooling from 135 K at 0.07 K min$^{-1}$, the changes in lattice constants due to the phase transition to ice XV start at 130 K, consistent with the appearance of the additional Bragg peaks shown in Figure 2. The changes in *b* and *c* in particular are quite fast initially but they slow down significantly as the temperature decreases. For example, half of the expansion in *c* is achieved within the first five degrees of the phase transition but the expansion keeps on taking place over the next 25 degrees before it finally levels off at around 100 K. This observation is fully consistent with an earlier calorimetry study of the phase transition which showed a sharp exotherm followed by a long tail upon cooling.[28]

At 80 K, the relative differences between the lattice constants of ice VI and ice XV are –0.16, –0.45 and +0.91% for the *a*, *b* and *c* lattice constants, respectively. The extent of these differences will of course depend on the degree of hydrogen order in the final ice XV and it may be possible to obtain larger differences in the future. In any case, it is clear that the *c* lattice constant changes the most during the ice VI to ice XV phase transition which corresponds to an increase in the distance between the apex water molecules of the hexameric unit shown in Figure 1(b).

Furthermore, the changes in lattice constants show that cooling slowly from the disordered phase is more effective in achieving a more ordered ice XV than slowly heating the pressure-quenched sample. The pressure-quenched sample represents a high free-energy state with respect to ice XV at low temperatures. So, in principle, the heating experiment could have revealed new metastable hydrogen-ordered states in line with Ostwald's rule of stages. However, given that (i) the same new Bragg peaks appear upon heating and cooling, and that (ii) the lattice constants change in the same fashion upon ordering, the formation of metastable stages does not seem to take place. The 'transient' hydrogen ordering observed upon heating seems to follow the same mechanistic pathway as the hydrogen ordering upon slowly cooling the disordered phase.

In addition to the changes in the lattice constants, Figure 4 also shows the changes in unit cell volume upon heating and cooling. It can be seen that the hydrogen ordering of ice VI goes along with an overall expansion of the unit cell volume. At 80 K, $D_2O$ ice XV is 0.14% less dense than $D_2O$ ice VI. Again, this value could be larger if a more hydrogen-ordered ice XV could be prepared. The volume expansion during the ice VI to ice XV phase transition explains why it is easier to achieve hydrogen ordering of ice VI upon cooling at ambient pressure compared to slow-cooling under pressure.[20] With increasing pressure, the volume expansion will become more and more unfavourable.

It is intriguing to speculate that increasing the pressure could open up a different pathway for hydrogen ordering compared to the one observed at ambient pressure. To test this possibility, a DCl-doped $D_2O$ ice VI sample was slow-cooled at 0.5 K min$^{-1}$ from 250 K at 1.4 GPa and recovered under liquid nitrogen at ambient pressure. The powder pattern of this sample at 80 K is shown as pattern (3) in Figure 3. Apart from a very weak feature marked by the question mark in the inset in Figure 3, the diffraction data is consistent with fully hydrogen-disordered ice VI. Considering that a partially hydrogen ordered ice XV has been previously obtained upon slow-cooling at 0.9 GPa,[20] this finding illustrates that increasing the pressure to 1.4 GPa makes the hydrogen ordering in ice VI even more difficult. A different pathway for hydrogen ordering does therefore not seem to emerge under pressure, at least up to 1.4 GPa.

Having discussed the formation of ice XV on a qualitative basis, we now turn to describing the structure of ice XV. Our labelling of the atoms within the unit cell of ice XV is shown in Figure 5(a). Figure 5(b) shows the three different chiral *A*, *B* and *C*-type structures that are possible for the individual networks. In order to build up candidate structures for fully hydrogen-ordered ice VI each of the two networks has to be assigned to either an *A*, *B* or *C*-type structure. This results in what we call *homo*-network structures (*A-A*, *B-B*, *C-C*) as well as *hetero*-network structures (*A-B*, *B-C*, *C-A*). Using graph theory, it has been shown that 45 distinct structures arise from these combinations.[23, 24] These can also be constructed in a straight-forward manner as follows. One of the structures shown in Figure 5(b) is assigned to the first network and this network is kept fixed. A second network is then chosen which is moved and rotated so that its oxygen atoms are in agreement with the oxygen positions of the second network. The 4-fold rotoinversion axis is then applied to the second network only which produces four structures. Four additional structures are obtained if the rotoinversion is performed onto one of the mirror images of the second network. The eight resulting structures are shown in Figure 5(c) for the case of *C-C* structures. The possible space group symmetries that arise from the symmetry relationships between the two networks are *P*1, *P*-1, *P*2$_1$, *Pn* and *Cc*. It is emphasised that the structures shown in Figure 5(c) are primitive unit cells which do not necessarily agree with the conventional crystallographic representations. The standard representation of the *Cc* unit cell, for example, has twice the volume of the primitive cell. The space groups containing the mirror images of the networks are *P*-1, *Pn* and *Cc*, whereas the *P*2$_1$ and *P*1 structures contain identical networks. Furthermore, in case of the *C-C* structures it is interesting to point out that there are two sets of four structures each which display the same the relative orientations of the apex molecules in the two networks. In structures 1-4 the two apex molecules of the two different networks appear parallel in the projection shown in Figure 5(c) whereas they are at a right angle in structures 5-8. Also, there are always two structures which are related by a 180° rotation of the 2$^{nd}$ network (1 and 4, 2 and 3, 5 and 8, as well as 6 and 7).

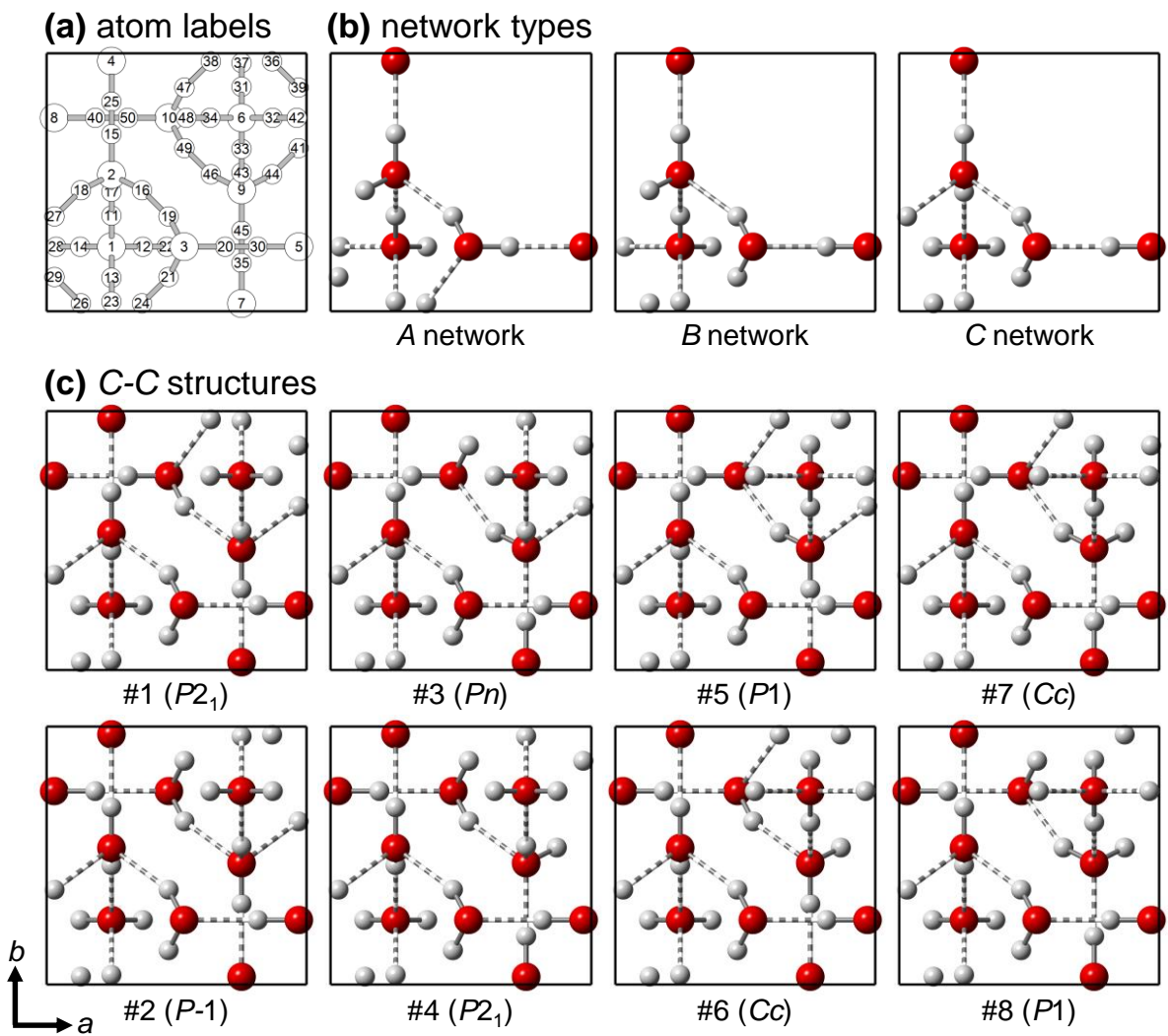

*Fig. 5* (a) Labels of the oxygen and hydrogen atoms in the two networks of a P1 structure. (b) Structures of A, B and C-type hydrogen-ordered networks. (c) The primitive unit cells of all possible structures containing two C-type networks.

The same space groups are obtained for the other two sets of *homo*-network structures, *A-A* and *B-B*, whereas the *hetero*-network structures all have *P*1 symmetry, *i.e.* there is no symmetry relationship between the networks. The labelling of the atoms of the 2nd network as shown in Figure 5(a) is only valid for *P*1 structures. In the higher symmetry space groups, the asymmetric unit contains only one network and the atoms of the second network therefore have the same labels as in the first network. The exact location of a counterpart atom in the second network of course depends on the details of the symmetry of the space group in question.

Each combination of network types gives eight possible structures which means 48 structures over all since there are six possibilities of combining the network types. Yet, in case of the *homo*-network structures, the two *P*1 structures are identical, and related by swapping the *a* and *b* axes. So, finally, there are 3×7 *homo*-network and 3×8 *hetero*-network structures which gives the 45 structures previously deduced using graph invariants.[23, 24]

Following our approach in ref. 20, the 21 structural models of the *homo*-network structures were refined against the neutron diffraction data collected after slowly cooling at ambient pressure. In general, there is obviously an inherent problem in refining structural models that are incompatible with the experimental data. So, as a first test of the structural models, only the lattice constants, scale factor, background function and peak profile parameters of the various fully hydrogen-ordered candidate structures were refined. The $\chi^2$ values in Figure 6 reflect the qualities of the fits to the experimental data. The structures with *Cc* space group gave some of the worst fits. This can be rationalised, for example, due to the fact that the 003 reflection, which is characteristic for ice XV, is not allowed in this space group. The *C-C* structures with *P*-1, *P*2$_1$ and *Pn* space groups were found to give the best fits. Interestingly, they are the members of one of the two sets of four *C-C* structures which have the same relative orientations of the apex molecules in the two networks. These structures were then taken forward for full structural refinements which included the occupancies of the hydrogen sites, the atomic coordinates of sites with fractional occupancies greater than 0.4 and the thermal displacement parameters. The $\chi^2$ values of the fits to the neutron diffraction data after slowly cooling at ambient pressure are shown in Figure 6.

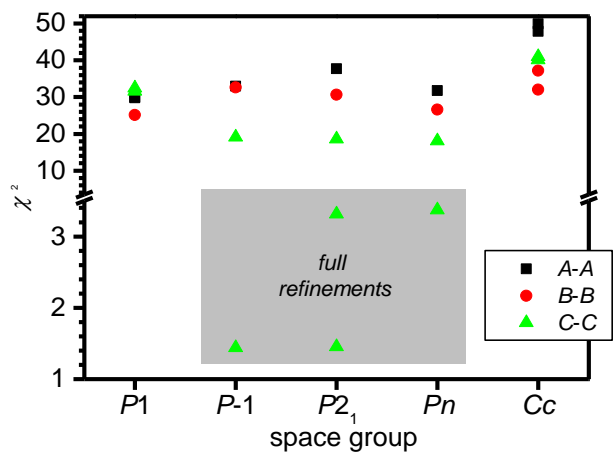

***Fig. 6*** *$\chi^2$ values of the Rietveld fits of the various homo-network candidate structures to the experimental neutron diffraction data after slowly cooling at ambient pressure (cf. pattern (2) in Figure 3). The shaded region indicates the results from full Rietveld refinements as described in the text.*

The *P*-1 structure containing two *C*-type networks provides the best fit to the experimental diffraction data. The fit of the *P*-1 model to the experimental diffraction data is shown as pattern (2) in Figure 3 and the crystal structure data is given in Table 1. Figure 7(a) shows the unit cell of the *P*-1 structures obtained from the Rietveld refinements. The occupancies of the hydrogen sites are indicated by a greyscale with white representing fully occupied and black fully unoccupied. The apex water molecules in ice XV are the most highly ordered, whereas the hydrogen bonds linking the hexameric units in *a* and *b* direction are the most disordered. The average absolute deviation from ½ is 0.28 which illustrates that the sample still contains significant amounts of hydrogen disorder.

Consistent with our previous study, one of the *P*2$_1$ structures corresponding to structure #1 in Figure 5(c) gives an almost equally good fit as the *P*-1 model.[20] As shown in Figure 7(a, b), the *P*-1 and *P*2$_1$ models describe, apart from slightly different distortions, essentially the same structure, which explains the similar $\chi^2$ values. The *P*2$_1$ model is therefore very close to nonpolar and it makes sense to use the antiferroelectric *P*-1 model to describe the structure of ice XV. Furthermore, as mentioned in ref. 20, the *P*-1 and *P*2$_1$ structures can be regarded as the internal and external racemates, respectively, which seem to be difficult to distinguish by diffraction. $P2_1$ implies a chiral crystal structure which means that different enantiomeric domains would have to exist. The interface between two chiral domains would require an unfavourable breaking of the ice rules.

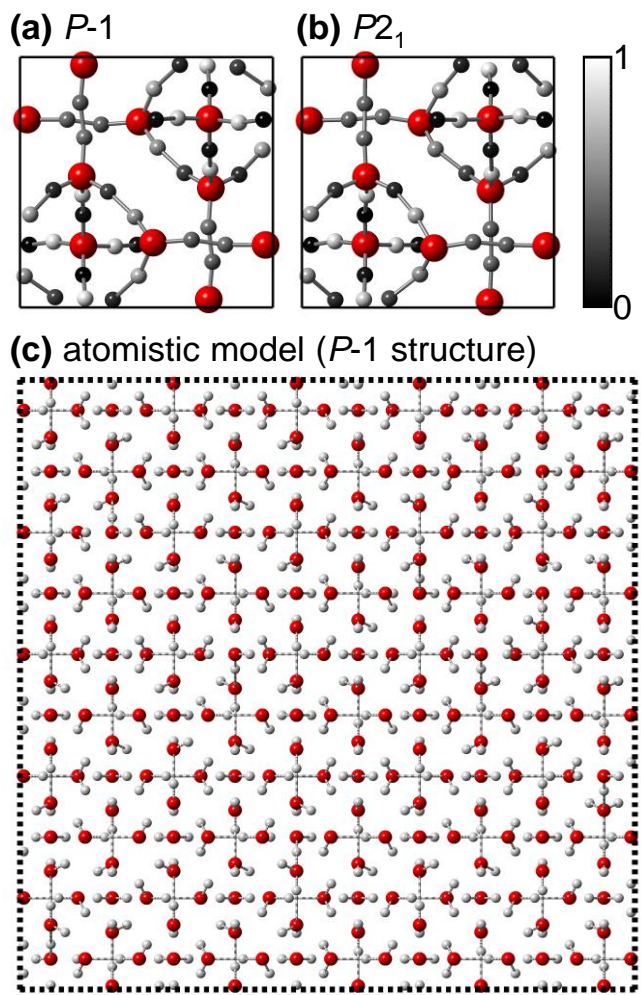

*Fig. 7* Crystal structures of the (a) P-1 and (b) $P2_1$ models of ice XV. The occupancies of the hydrogen sites are indicated by the greyscale. (c) Atomistic model of a 5x5x6 supercell consistent with the P-1 structure of ice XV. The maximal (average) deviation from the target occupancies shown in Table 1 is 0.087 (0.046). To prevent atomic overlaps, only atoms in the $0 < z < 1/6$ range are shown. Covalent O-H bonds are indicated by solid and O⋯H hydrogen bonds by dashed lines.

Experimentally, the lattice constants have been shown to behave in the following manner upon going from ice VI to ice XV: (i) Both *a* and *b* contract, (ii) if the structure of ice XV is defined as shown in Figure 5 then *b* changes more than *a*, and (iii) the most pronounced change is the expansion in *c*. Figure 8 shows the ratios of the lattice constants from DFT calculations of all possible *homo*-network structures with respect to the lattice constants of a fully hydrogen-disordered 3x3x3 supercell. Considering that there is strong evidence for *C*-type networks in ice XV, the *P*-1 model, labelled as 2C in Figure 8, is the only model that fulfils all of the conditions giving further support for this structural model of ice XV. It can also be seen from Figure 8 that the *homo*-network *P*1 structures are indeed related to one another by swapping the *a* and *b* axes.

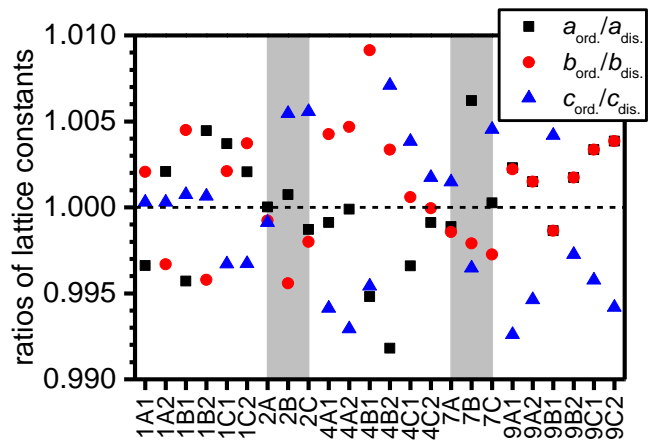

*Fig. 8* Ratios of the lattice constants obtained from DFT calculations of the various homo-network structures with respect to the lattice constants of a fully hydrogen-disordered 3x3x3 supercell. The first number of the labels of the structures is the number of the space group and followed by the type of networks the structure contains. For some combinations there are two distinct structures for a given space group and network type.

Figure 7(c) shows a slice of a supercell consistent with the average structure of the *P*-1 model obtained from Rietveld refinements. The average percentage of *C*-type motifs calculated from all six slices of the 5x5x6 supercell is 45.3 ± 4.4 %, and 82.0 ± 0.0 % of the apex molecules, including those from the *C*-type motifs, appear as 'straight' viewed along *c*. This illustrates that the partial hydrogen disorder within ice XV causes the appearance of structural motifs that are outside of the *A*, *B* and *C* classification. Taken to the extreme, the fully hydrogen-disordered structure shown in Figure 1(c), for example, contains only 10.7 ± 6.4 % *C*-type motifs.

Komatsu and co-workers recently suggested that ice XV is best described with *Pmmn* space group symmetry and that the local structure can be described as a mixture of *C-C* motifs with different symmetry relationships between the hexameric units.[30] The first step of our preparation procedure is the rapid freezing of the DCl solution inside a piston-cylinder setup at liquid nitrogen temperature. In contrast to this, the procedure in ref. 30 involved loading of the liquid solution into the pressure setup at room temperature followed by slowly freezing the solution at ambient pressure. Firstly, hydrochloric acid is a very aggressive acid that reacts very readily with metals leading to a reduction of its concentration and the formation of metal chlorides. But furthermore, the rapid freezing of the liquid solution in our studies could be important with respect to achieving the highest possible concentrations of hydrochloric acid in the ice. Following the formation of ice VI under pressure, the preparation of ice XV was also achieved by slow-cooling at ambient pressure in ref. 30. However, the maximal intensity of the 003 Bragg peak in the diffraction data shown in ref. 30 is only less than half that observed by us. This implies that the ice XV prepared in ref. 30 is considerably less ordered than achieved by us.

The *Pmmn* model suggested by Komatsu and co-workers was refined against our experimental data in a next step. The hydrogen order in the *Pmmn* model is described by a single order parameter and it requires the hydrogen sites of the hydrogen bonds that connect the hexameric units in *a* and *b* direction to be fully hydrogen disordered. So in this respect, the *Pmmn* model captures a structural trend shown in Figure 7(a). However, unrealistic atomic movements were observed during the refinement of the *Pmmn* model even if the refinement of the atomic coordinates was heavily damped. The resulting structure showed no resemblance to the ice VI framework, the molecular characters of the water molecules were lost entirely and unphysical interatomic distances were found. While it seems possible that the *Pmmn* model gives reasonable fits to less-ordered ice XV, it certainly cannot be used to describe the structure of the ice XV made in this study. It is also noted that the position of the 003 peak of the *Pmmn* model does not accurately reproduce the position of the experimental peak shown in Fig. 6 in ref. 30. Considering that using the high-symmetry $P4_2/nmc$ space group to describe the structure of the hydrogen-disordered ice VI is a crystallographic 'trick' which does not reflect the local structure, we argue that only structural models that allow full hydrogen order should be used to describe the average structure of ice XV independent of the actual extent of hydrogen order.

**Conclusions**

To conclude, we have described several aspects of the ice VI to ice XV hydrogen ordering phase transition. The expansion of the unit cell volume upon hydrogen ordering explains why the ordering process is fastest at ambient pressure and slows down significantly if carried out under pressure. Consistent with an earlier calorimetric study,[28] the phase transition takes place over a quite large temperature range and it is very difficult to achieve high degrees of hydrogen order. There is a strong tendency for *C*-type networks to form during the hydrogen order phase transition and the most ordered structure obtained so far is best described with *P*-1 space group symmetry. The *Pmmn* model reflects some of the characteristics of the hydrogen order in ice XV but it is overall incompatible with our diffraction data. Future experimental work should focus on obtaining more highly ordered ice XV.

**Acknowledgements**

We thank the Royal Society for a University Research Fellowship (CGS, UF100144), the Leverhulme Trust for a Research Grant (RPG-2014-04), ISIS for granting access to the POLARIS instrument and Dr R. Smith for help with the neutron diffraction measurements.

**Table 1.** Fractional atomic coordinates and occupancies for DCl-doped $D_2O$ $P$-1 ice XV obtained after cooling at 0.07 K min$^{-1}$ at ambient pressure. The isotropic thermal displacement parameters multiplied by 100 are 0.982(8) and 1.611(7) for oxygen and deuterium atoms, respectively. The lattice constants are: $a$ = 6.23416(6) Å, $b$ = 6.21697(7) Å, $c$ = 5.83377(6) Å, $\alpha$ = 90.033(5) °, $\beta$ = 89.9219(30) ° and $\gamma$ = 89.999(6) °.

| atom | x | y | z | occupancy |
| --- | --- | --- | --- | --- |
| O1 | 0.2495(16) | 0.2469(14) | 0.2658(4) | 1 |
| O2 | 0.2441(14) | 0.5227(5) | 0.6413(5) | 1 |
| O3 | 0.5269(5) | 0.2592(13) | 0.8817(5) | 1 |
| O4 | 0.2524(16) | 0.9675(5) | 0.6419(5) | 1 |
| O5 | 0.9720(5) | 0.2478(14) | 0.8850(4) | 1 |
| D11 | 0.2500 | 0.3704 | 0.3635 | 0.086(2) |
| D12 | 0.3755(4) | 0.2411(12) | 0.1665(5) | 0.919(2) |
| D13 | 0.2500 | 0.1296 | 0.3635 | 0.019(2) |
| D14 | 0.1323(4) | 0.2590(11) | 0.1537(5) | 0.973(2) |
| D15 | 0.2537(28) | 0.6790(7) | 0.6316(10) | 0.531(4) |
| D16 | 0.348824 | 0.462806 | 0.73248 | 0.319(4) |
| D17 | 0.2470(19) | 0.4490(4) | 0.4937(7) | 0.914(2) |
| D18 | 0.1325 | 0.4693 | 0.7155 | 0.219(4) |
| D19 | 0.4587(9) | 0.3610(9) | 0.7836(9) | 0.681(4) |
| D20 | 0.6839(7) | 0.2598(23) | 0.8833(11) | 0.488(5) |
| D21 | 0.4720(8) | 0.1274(8) | 0.8065(9) | 0.770(3) |
| D22 | 0.4629 | 0.2500 | 0.0138 | 0.081(2) |
| D23 | 0.2576(13) | 0.0348(4) | 0.4936(6) | 0.981(2) |
| D24 | 0.3675 | 0.0307 | 0.7155 | 0.230(3) |
| D25 | 0.2375(19) | 0.8175(8) | 0.6359(12) | 0.469(4) |
| D26 | 0.138843 | 0.04094 | 0.744109 | 0.308(5) |
| D27 | 0.0447(7) | 0.3722(8) | 0.8175(9) | 0.781(4) |
| D28 | 0.0371 | 0.2500 | 0.0138 | 0.027(2) |
| D29 | 0.0200(9) | 0.1239(9) | 0.8087(11) | 0.692(5) |
| D30 | 0.8167(7) | 0.2681(15) | 0.8895(10) | 0.512(5) |